\documentclass[final,5p,times,twocolumn]{elsarticle}

\usepackage{amssymb}
\usepackage{amsmath}
\usepackage{amsfonts}
\usepackage{amsthm}
\usepackage{xcolor}
\usepackage{graphicx}
\usepackage{float}
\usepackage{color,soul}

\usepackage{hyperref}
\hypersetup{
    colorlinks=true, %set true if you want colored links
    linktoc=all,     %set to all if you want both sections and subsections linked
    citecolor=black,
    filecolor=black,
    linkcolor=black,
    urlcolor=black
}
\newcounter{bla}

\journal{Computer Physics Communications}

\DeclareUnicodeCharacter{2060}{}

\begin{document}

\begin{frontmatter}

\title{FRET$-$Calc: A Free Software and Web Server for F\"{o}rster Resonance Energy Transfer Calculation}

\author[a,b]{Leandro Benatto\corref{author}}
\author[a]{Omar Mesquita}
\author[b]{Jo\~{a}o L. B. Rosa}
\author[b]{Lucimara S. Roman}
\author[b]{Marlus Koehler}
\author[a,c]{Rodrigo B. Capaz}
\author[a]{Grazi\^{a}ni Candiotto\corref{author}}

\address[a]{Institute of  Physics, Federal University of Rio de Janeiro, 21941$-$909, Rio de Janeiro$-$RJ, Brazil.}
\address[b]{Department of Physics, Federal University of Paran\'{a}, 81531$-$980, Curitiba$-$PR, Brazil.}
\address[c]{Brazilian Nanotechnology National Laboratory (LNNano), Brazilian Center for Research in Energy and Materials (CNPEM), 13083$-$100, Campinas$-$SP, Brazil.}

\cortext[author]{Corresponding authors: \\ lb08@fisica.ufpr.br and gcandiotto@iq.ufrj.br\\ DOI: \href{https://doi.org/10.1016/j.cpc.2023.108715}{10.1016/j.cpc.2023.108715}}

\begin{abstract}
F\"{o}rster Resonance Energy Transfer Calculator (FRET$-$Calc) is a program and web server that analyzes molar extinction coefficient of the acceptor, emission spectrum of the donor, and the refractive index spectrum of the donor/acceptor blend. Its main function is to obtain important parameters of the FRET process from experimental data, such as: (i) effective refractive index, (ii) overlap integral, (iii) F\"{o}rster radius, (iii) FRET efficiency and (iv) FRET rate. FRET$-$Calc is license free  software that can be run  \textit{via} dedicated web server (\href{https://nanocalc.org/}{nanocalc.org})  or downloading the program executables (for \href{https://github.com/NanoCalc/FRETCalc/releases/download/FRETCalc-1.0-alpha/FRETCalc-Unix.tar.gz}{\textit{Unix}}, \href{https://github.com/NanoCalc/FRETCalc/releases/download/FRETCalc-1.0-alpha/FRETCalc-Windows.tar.gz}{\textit{Windows}}, and \href{https://github.com/NanoCalc/FRETCalc/releases/download/FRETCalc-1.0-alpha/FRETCalc-MacOS.tar.gz}{\textit{macOS})} from the \href{https://github.com/NanoCalc/FRETCalc}{FRET$-$Calc} repository on \textit{GitHub}. The program features a user$-$friendly interface, making it suitable for materials research and teaching purposes. In addition, the program is optimized to run on normal computers and is lightweight. An example will be given with the step by step of its use and results obtained.
\end{abstract}


\end{frontmatter}

%\linenumbers

\section{Introduction}
Fluorescence (or F\"{o}rster) resonance energy transfer is physical process based on the long range dipole$-$dipole interactions between molecules that are fluorescent \cite{mikhnenko2015}.⁠ In the FRET process, upon photofexcitation, the excited donor (D) molecule can transfers its excess energy to a nearby acceptor (A) molecule. Once the process is completed, the donor returns to the ground state and the acceptor reaches the excited state. In the FRET model, the FRET rate ($k_{FRET}$) between donor and acceptor depends on the F\"{o}rster radius, the D$-$A distance, and the exciton lifetime of the donor in the absence of the acceptor \cite{hildebrandt2013}.⁠ FRET usually occurs for D$-$A distances between 10 to 100 \AA, a range that is comparable to the dimensions of most biological macromolecules \cite{hillisch2001}.⁠ For lower distances, known as contact zone, the ideal dipole approximation of F\"{o}rster theory breaks down and the energy transfer process occurs predominantly from Dexter mechanism \cite{van2013forster}.

The FRET phenomena plays a significant role in many technological applications such as organic photovoltaics\cite{benatto2021,du2020,karki2020} organic light$-$emitting diodes\cite{heimel2018,braveenth2021}⁠ and biosensing\cite{chou2015,tian2021} among others \cite{alhadrami2021,andre2021,sharma2021}⁠. For example, the optimization of FRET parameters is important to increase the exciton diffusion length of organic materials \cite{firdaus2020,candiotto2017}⁠. Additionally, measuring the competition between FRET rate and charge transfer rate is critical to understanding charge dynamics in different devices \cite{xu2022,xu2020}.⁠ Considering FRET$-$based biosensors, the high sensitivity, specificity, fast responsibility, and simplicity enables its application in medical diagnosis, food safety, and environmental monitoring \cite{zhang2019,melnychuk2018}.⁠ Therefore, software that calculates FRET parameters can be useful in different areas and will allow obtaining relevant information at the molecular level, increasing the understanding of the photophysical process in materials with strong potential for technological applications.⁠

Despite the procedure to calculate F\"{o}rster radius is well known in the literature, it is not so simple to be implemented \cite{benatto2021,karuthedath2021,candiotto2020}. Currently there are some softwares which calculate the FRET parameters. However, these softwares are commonly implemented to obtain the parameters from theoretical quantum chemical calculation \cite{kosenkov2016,kosenkov2022}, single molecule \cite{ingargiola2016} or from the analysis of molecular structure of biomolecules, such nucleic acid (RNA and DNA) \cite{preus2012}. Thus, softwares that obtain the FRET parameters directly from experimental data and that can be applied to study organic materials such as polymers, fullerenes, non-fullerenes and dyes are lacking. Motivated by this deficiency in the literature, we developed a free software and web server, F\"{o}rster Resonance Energy Transfer Calculator (FRET$-$Calc), that is easy to handle with a simple and intuitive graphic interface for the FRET rate acquisition. In addition to applicability in materials research, the intuitive platform of the software and web server enable them to be used for didactic purposes.

FRET$-$Calc was developed in Python programming language and the main features are to analyze spectral data of the acceptor (molar extinction coefficient), donor (emission), and donor/acceptor blend (refractive index). After processing the data, the main parameters of the FRET process are obtained, such as: effective refractive index, overlap integral, F\"{o}rster radius, FRET efficiency and FRET rate.

\section{Theoretical Background}
The excitation energy from a D molecule is transferred to the A by resonance coupling of the donor dipole and acceptor dipole. Thus, in FRET model, the rate of energy transferred, $k_{FRET}$, between donor and acceptor is given by Ref. \citep{van2013forster}

\begin{equation}\label{eq-kfret}
{k_{FRET}=\dfrac{1}{\tau_{D}}\left(\dfrac{R_{0}}{R}\right)^{6},}
\end{equation}

\noindent where $\tau_{D}$ is the exciton lifetime of the donor in the absence of the acceptor, $R_{0}$ is the F\"{o}rster radius, and $R$ is the donor$-$acceptor distance. The FRET efficiency is traditionally described as

\begin{equation}
\eta_{FRET}=\dfrac{R_{0}^{6}}{R_{0}^{6}+R^{6}}.
\end{equation}

\noindent In the case of multiple donors and acceptors, $\eta_{FRET}$ changes for

\begin{equation}
\eta_{FRET}=\dfrac{n_{A} R_{0}^{6}}{n_{A} R_{0}^{6}+R^{6}},
\end{equation}

\noindent where $n_{A}$ is the number of acceptors per donor \cite{chou2015,clapp2004}.⁠ This simple equation was derived considering equal distances between all donors and acceptors and assuming that all FRET rates are equal for any D$–$A FRET pair \cite{hildebrandt2013}.⁠ As the number of acceptors per donor increases, the FRET probability becomes higher.

In nanometer, $R_{0}$ can be expressed as 

\begin{equation}\label{eq-r0}
R_{0} = 0.0211\left(\dfrac{\kappa^{2}\Theta_{D}J(\lambda)}{n^{4}}\right)^{1/6},
\end{equation}

\noindent where kappa$-$squared $\kappa^{2}$ is the dipole$-$dipole orientation factor, $n$ is the refractive index of the medium, $\Theta_{D}$ is the quantum yield of the donor fluorescence in the absence of the acceptor, and $J$ is the spectral overlap integral between the acceptor absorption spectrum and the area$-$normalized emission spectrum of the donor. Through Eq. \ref{eq-r0} it is possible to conclude that a higher donor quantum yield and larger overlap of donor emission and acceptor$-$excitation spectra are fundamental keys to achieve greater energy transfer, which will result in a better FRET signal.

The FRET mechanism is based on the coupling of the transition dipoles of the emission donor and the acceptor excitation. Thus, for energy transfer to occur with varying efficiencies, the dipoles must have a favorable orientation relative to each other. The kappa$-$squared can assume values from $0$ to $4$, according to the following conditions: 1) $\kappa^{2} = 0$, when the electric field produced by the donor’s emission is oriented perpendicularly to the acceptor’s dipole transition. In this configuration there is no energy transfer by the FRET mechanism; 2) $\kappa^{2} = 1$, when the D and A dipoles are in parallel position to each other and perpendicular to the bond vector; 3) $\kappa^{2} = 4$, when the dipoles are aligned (\textit{e.g.}, their dipole moments are in a parallel position to each other and to the bond vector); 4) when the dipoles have a random orientation, but fixed in time (example, solid state film), the value of $\kappa^{2} = 0.476$ is usually assumed; 5) In the case of dipoles having random orientation, but rotating freely in time (example, in solution), the value of $\kappa^{2} = 2/3$ is usually assumed. There is an extensive debate in the literature about the general validity of those assumptions. The interested reader is referred to Ref. \cite{vandermeer2020}⁠ for further discussions.

In the derivation FRET mechanism, the overlap integral $J$ included in transfer rate equation appears to satisfy the dipole energy resonance condition, \textit{i.e.} ensure energy conservation during transfer \cite{forster1948,valeur2012}. The probability per unit time of the spectroscopic transitions of a molecule depends on the distribution of its electronic energy levels in the ground and excited states. This distribution can be obtained by measuring the emission and/or absorption spectra. Consequently, $J$ is proportional to the overlap between the emission spectra of the donor and the absorption spectra of the acceptor.

If the materials' refractive index is not available, the Kramers$-$Kronig relation can be used following the procedure described in detail in Ref. \cite{balawi2020}⁠. For the procedure, the absorption coefficient of each blend is required. Karuthedath \textit{et al.} \cite{karuthedath2021}⁠ argued that the Kramers$-$Kronig relation approximates the refractive indexes very well, with a few percentage of deviation in relation to ellipsometry measurements. From the $n(\lambda)$ information, the effective refractive index $n_{eff}$ can be computed as a weighted for the amplitude of the overlap between the donor’s emission and the acceptor’s absorption \cite{karuthedath2021}.

For calculating $J$, the acceptor absorption spectrum and the emission spectrum of the donor is necessary. Thus, $J(\lambda)$ can be calculated in units of mol$^{–1}$ L cm$^{-1}$ nm$^{-4}$ using the formula \cite{karuthedath2021}

\begin{equation}\label{eq-J}
J(\lambda) = \int^{\infty}_{0} \varepsilon_{A}(\lambda)\lambda^{4}F_{D}(\lambda)d\lambda,
\end{equation}

\noindent where $\varepsilon_{A}$ is the molar extinction (attenuation) coefficient of the acceptor and $F_{D}(\lambda)$ is the wavelength-dependent donor emission spectrum normalized to its area \cite{chou2015}.⁠ If the absorption measurements were done in solution, the extinction coefficient spectrum can be obtained using Lambert$-$Beer’s law

\begin{equation}
\varepsilon_{A}(\lambda) =\dfrac{A(\lambda)}{lc},
\end{equation}

\noindent where $A(\lambda)$ is the background-corrected absorbance spectrum, $c$ is the concentration of the sample and $l$ is the light path length \cite{hildebrandt2013}.⁠ If the absorption measurements were done in thin films, can be obtained using the films’ absorption coefficients $\alpha(\lambda)$ (units cm$^{-1}$), molecular weight $M_{W}$, and the density $d$ of materials \cite{karuthedath2021},

\begin{equation}
\varepsilon_{A}(\lambda) =\dfrac{\alpha(\lambda)M_{W}}{d}.
\end{equation}

\noindent For the density, a typical value for polymeric materials is $1.2$ kg/L \cite{density2022}.⁠ The molecular weight of the acceptor in unity of g/mol is a parameter that is generally easy to find or calculate. Therefore, $\varepsilon_{A}(\lambda)$ is expressed in units of cm$^{-1}$ mol$^{-1}$ L or M$^{-1}$ cm$^{-1}$.

\section{Software architecture, implementation and requirements}
\begin{figure}[H]
\centering
    \includegraphics[width=\linewidth]{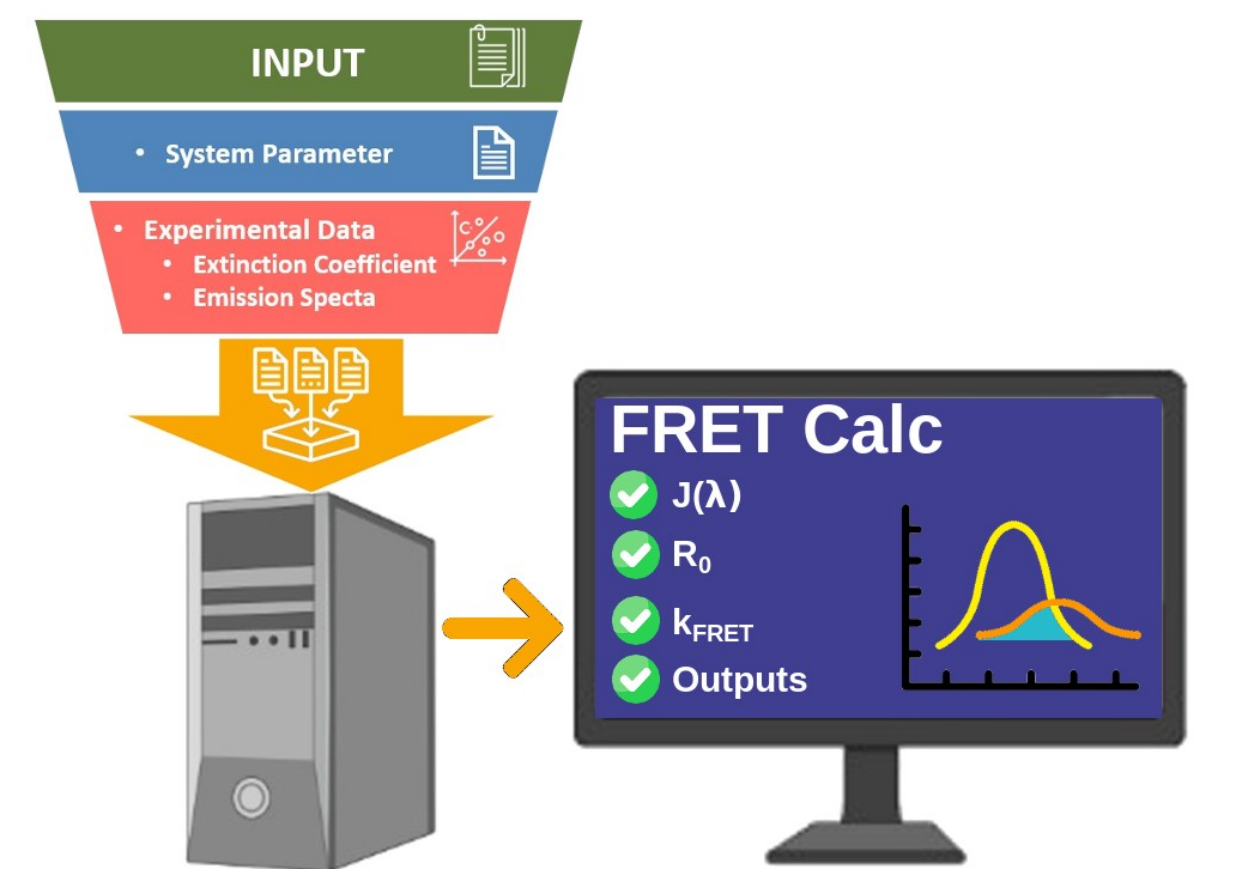}
    \caption{Graphical representation of FRET$-$Calc workflow.}
    \label{fig-panel-workflow}
\end{figure}

The FRET$-$Calc is a license free code and can be used \textit{via} a dedicated web server or downloading the binary files. The web server can be accessed in \href{https://nanocalc.org/}{nanocalc.org} and it is compatible with the main browsers like \textit{Chrome}, \textit{Firefox}, \textit{Safari}, \textit{Opera}, \textit{Brave}, \textit{Edge} and \textit{Internet Explorer}. The binary files for \href{https://github.com/NanoCalc/FRETCalc/releases/download/FRETCalc-1.0-alpha/FRETCalc-Unix.tar.gz}{\textit{Unix}}, \href{https://github.com/NanoCalc/FRETCalc/releases/download/FRETCalc-1.0-alpha/FRETCalc-Windows.tar.gz}{\textit{Windows}}, and \href{https://github.com/NanoCalc/FRETCalc/releases/download/FRETCalc-1.0-alpha/FRETCalc-MacOS.tar.gz}{\textit{macOS}} operational systems are available for download at \href{https://github.com/NanoCalc/FRETCalc}{FRET$-$Calc} repository on \textit{GitHub}. FRET$-$Calc is composed by four modules that is shown in workflow of Figure \ref{fig-panel-workflow}. In Module 1) system parameters  and experimental data for extinction coefficient (acceptor), emission spectra (donor) and refractive index (donor$-$acceptor) are read; 2) calculation of overlap integral ($J(\lambda)$); 3) determination of F\"{o}rster Radius ($R_{0}$) and 4) determination of FRET rate ($k_{FRET}$). The program is implemented in Python 3 (v. 3.6) \cite{van2009} and makes use of four Python libraries, namely \textit{Pandas} \cite{pandas2010}, \textit{NumPy} \cite{harris2020},⁠ and \textit{SciPy} \cite{virtanen2020}⁠ for data manipulation and \textit{Matplotlib} \cite{hunter2007}⁠ for data visualization. The software was designed to be very simple to use and take up little disk space, around 85 MB.

\section{Program and Application}
\begin{figure}[h!]
\centering
    \includegraphics[width=\linewidth]{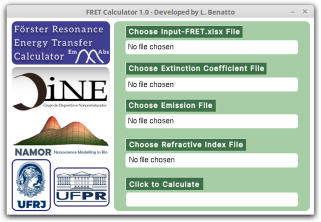}
    \caption{Program interface.}
    \label{fig-panel-interface}
\end{figure}

When the program runs (or web server is accessed) a graphical interface is loaded, as shown in Figure \ref{fig-panel-interface}. It indicates that it is necessary to enter information about the system to be analyzed. The first request is for an excel file with some information about the donor and acceptor materials.

\begin{figure}[h!]
\centering
    \includegraphics[width=\linewidth]{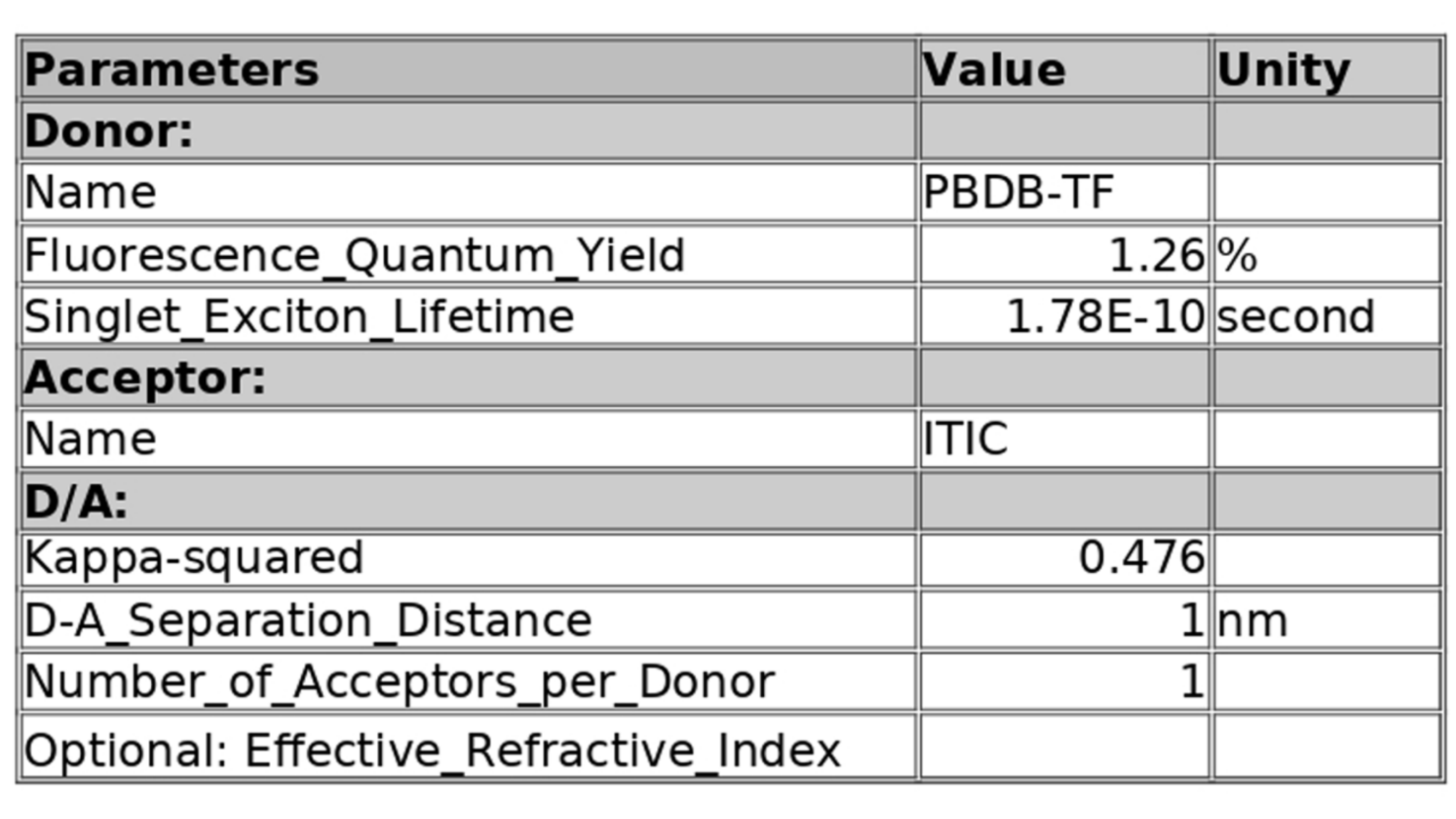}
    \caption{Input-FRET.xlsx file with the input parameters.}
    \label{fig-panel-input}
\end{figure}

An example of input file is shown in Figure \ref{fig-panel-input} and it is available for the user to fill in at \href{https://github.com/NanoCalc/FRETCalc}{FRET$-$Calc} repository. In the example, we use information regarding the PBDB$-$TF \cite{zhang2015}⁠ (also known as PM6 and PBDB$-$T$-$2F \cite{zheng2020}⁠) donor polymer and the ITIC \cite{zhao2016}⁠ molecular acceptor. For pristine PBDB$-$TF, the exciton lifetime is 178 ps and the quantum yield of fluorescence is 1.26\% \cite{karuthedath2021}.⁠ The dipole$-$dipole orientation factor was set equal to 0.476, corresponding to a rigid and random relative orientation of the molecular dipoles in a solid-state films (for randomized rotation diffusion of molecules, not being the solid-state, $\kappa^{2}$ = 2/3) \cite{firdaus2020,lin2014}.⁠ For the donor$-$acceptor separation distance, the standard value of 1 nm is generally used for organic materials in solid$-$state. There is also in the input file an option to insert the effective refractive index, $n_{eff}$, because it is common that the user does not have available the refractive index spectrum of the D/A blend.

\begin{figure}[h!]
\centering
    \includegraphics[width=\linewidth]{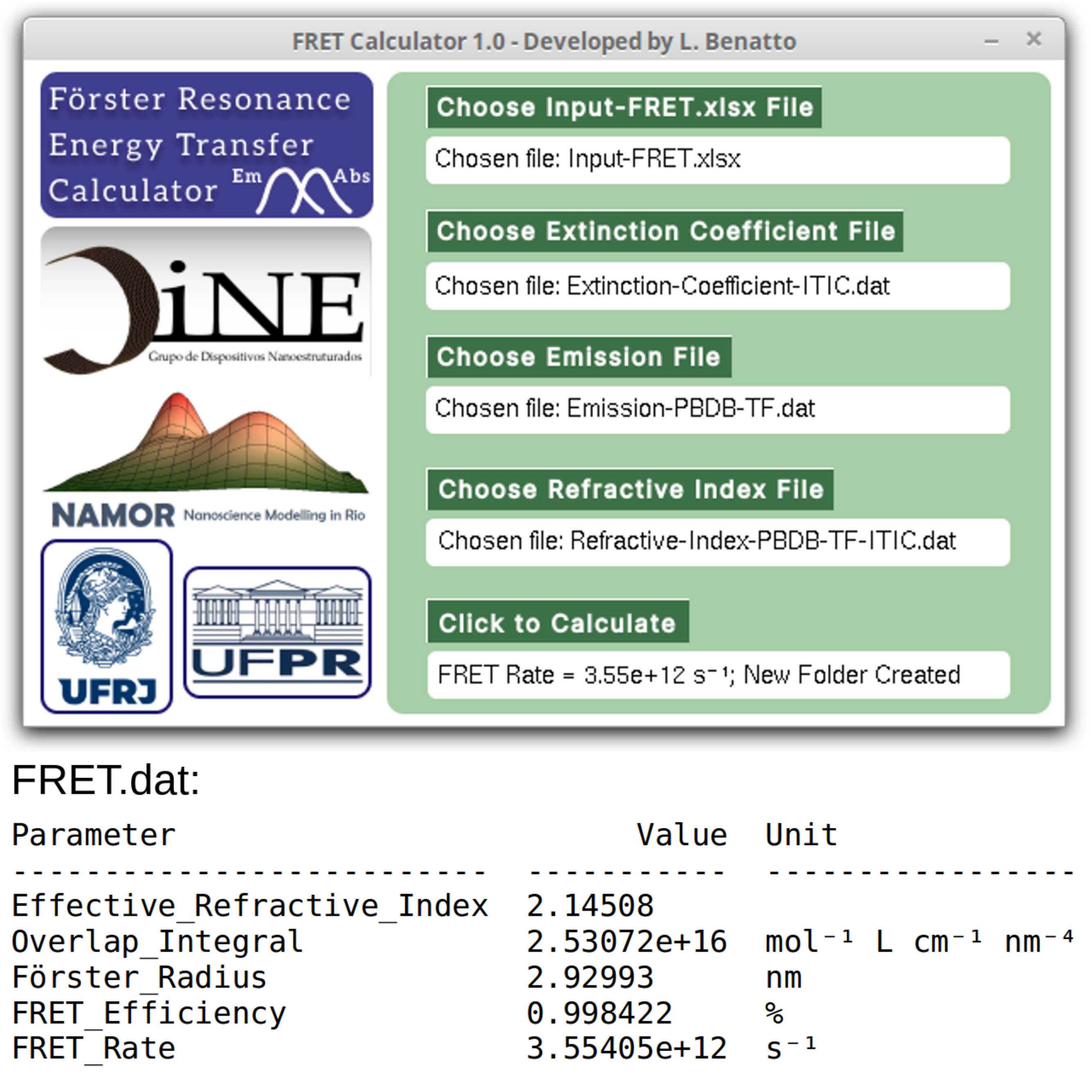}
    \caption{Provided all the necessary information, the program calculates five FRET properties of the system.}
    \label{fig-panel-output-dat}
\end{figure}

\begin{figure*}[t!]
\centering
    \includegraphics[width=\linewidth]{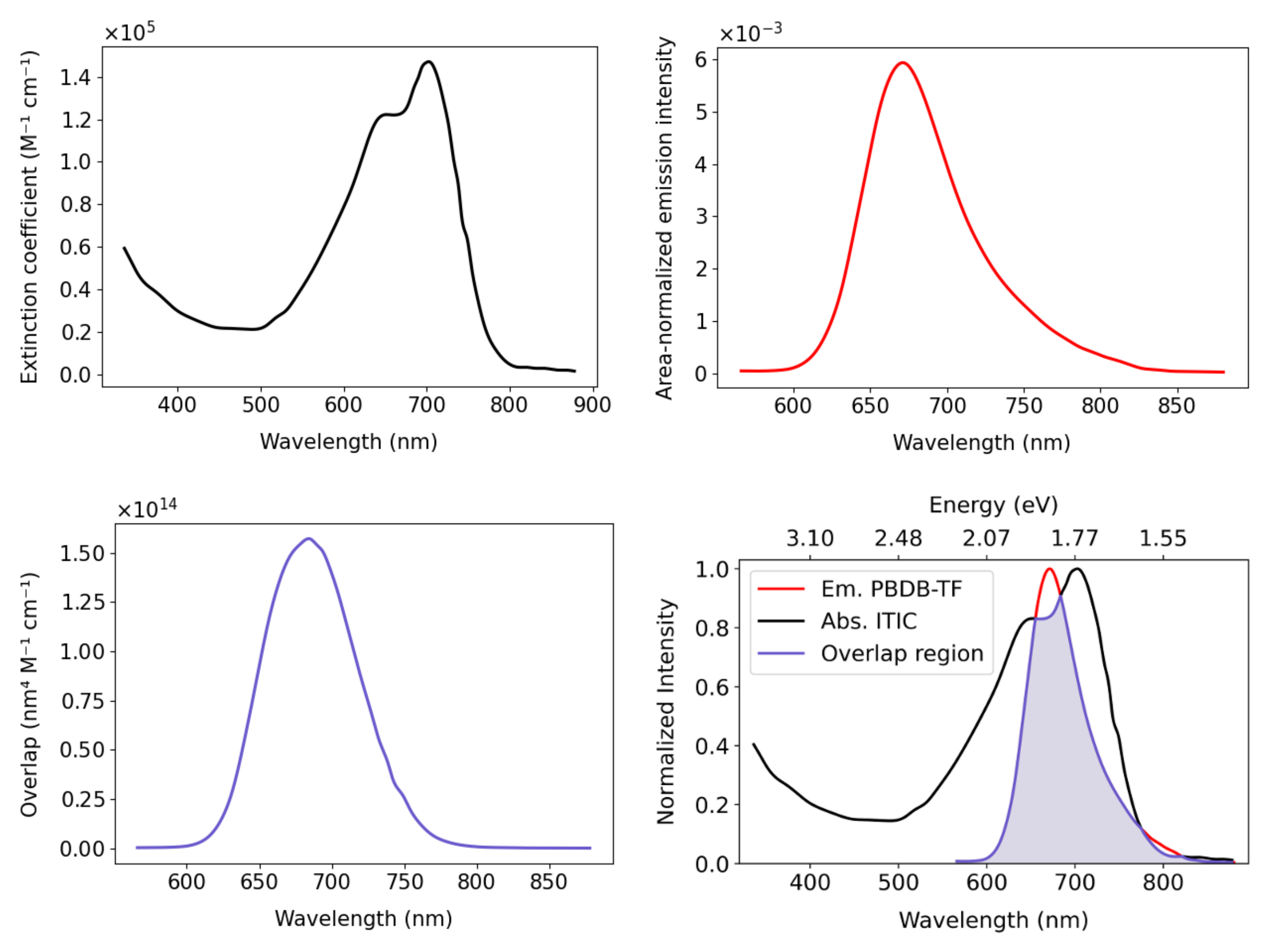}
    \caption{Acceptor’s extinction coefficient (top left), area$-$normalized donor emission spectrum (top right), overlap function (bottom left), and overlap region (highlighted blue region) between the PBDB$-$TF emission and ITIC absorption spectra (bottom right). Experimental results of acceptor’s extinction coefficient and donor emission spectrum of ref.\cite{yang2019}.}
    \label{fig-panel-output}
\end{figure*}

\begin{figure*}[t!]
\centering
    \includegraphics[width=\linewidth]{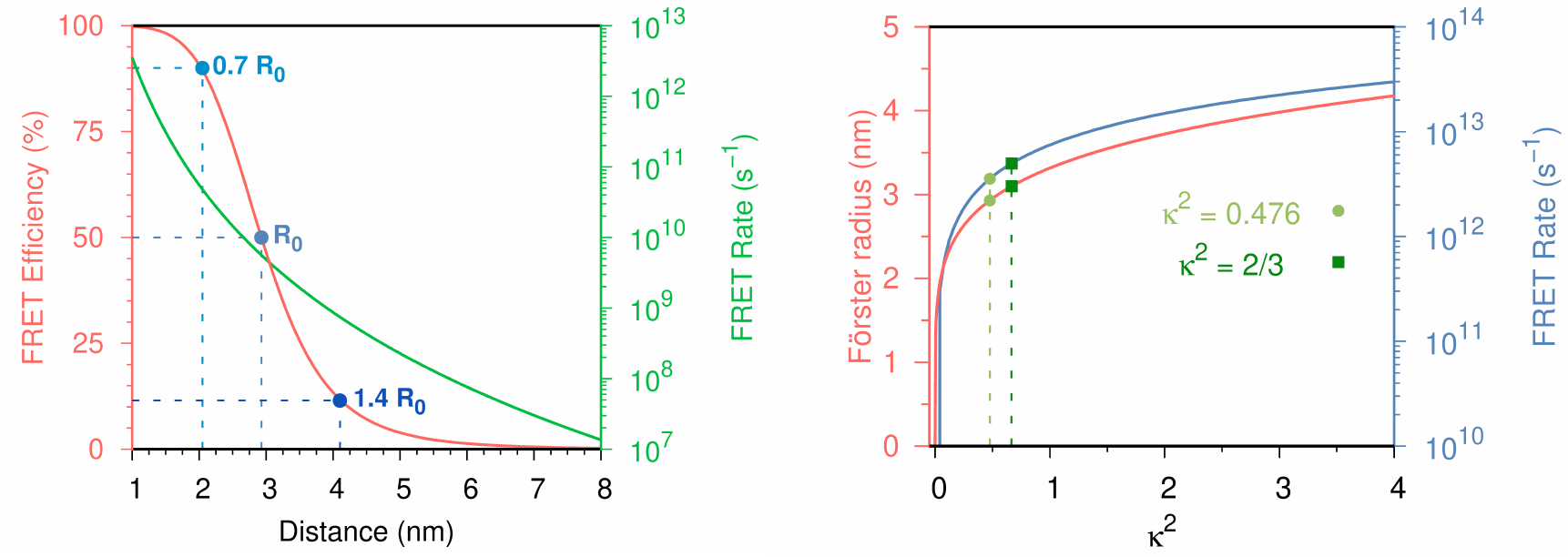}
    \caption{(Left) Behavior of FRET efficiency (red) and FRET rate in log scale (green) versus donor$-$acceptor separation distance. The dotted lines on the graph indicates the F\"{o}rster radius at which the efficiency is 90, 50 and 10\%. (Right) Dependence of F\"{o}rster radius (red) and FRET rate in log scale (blue) with kappa$-$squared. The dotted lines in the graph exemplify two values widely used in the literature for kappa$-$squared.}
    \label{fig-panel-fret-behaviour}
\end{figure*}

The next files that the program requests are the acceptor's extinction coefficient spectrum, the donor's emission (photoluminescence) spectrum and the refractive index of the D/A blend (if available). The donor's emission does not necessarily have to be normalized, as this procedure is performed by the program. The three input spectra should be provided as a function of wavelength, not energy. It is worth mentioning that the three spectra mentioned above do not necessarily have the same step size for the wavelength, because the program will perform a cubic interpolation of the data and analyze the results using the same interval. The data points also can be non-equidistant. In this way, it is possible to use spectra obtained from different sources. With all the information provided, the program will calculate the effective refractive index, overlap integral, F\"{o}rster radius, FRET efficiency and the FRET rate. This information is printed in the FRET.dat file, as can be seen in Figure \ref{fig-panel-output-dat}. In addition, four graphs will be generated (Figure \ref{fig-panel-output}), which are acceptor’s extinction coefficient (top left), area-normalized donor emission spectrum (top right) and overlap function (bottom left). A well$-$known graph in the literature that illustrates the overlap region will also be generated, see Figure \ref{fig-panel-output} (bottom right). All output files are saved in a new folder created automatically.

Just to confirm the validity of the method implemented here, the FRET parameters of some D$/$A blends available in the supporting information of Ref. \cite{karuthedath2021}⁠ were calculated using the same dataset. The results matched, and no relevant differences were observed.

To complement the program demonstration, in the Figure \ref{fig-panel-fret-behaviour} (left) we presented the typical behavior of FRET efficiency with donor$-$acceptor separation distance. Due to dependence on the sixth power of distance, the FRET Efficiency drops off very fast in the range of $R$ is 0.7$-$1.4$R_{0}$, corresponding to 90$-$10\% FRET efficiency. In the special case where $R = R_{0}$, FRET efficiency has a value of 50\%. Still in the Figure \ref{fig-panel-fret-behaviour} (left), the variation of FRET rate is presented. Note that the drop$-$in rate averages approximately one order of magnitude for each nanometer increased in the studied interval. In short, the donor$-$acceptor separation distance is a important parameter in the energy transfer process.

Remember that kappa$-$squared is a number between 0 and 4 and may have specific values depending on the orientation between the dipole moments of donor and acceptor \cite{vandermeer2020}.⁠ As an additional feature, the program also permits to easily access the influence of kappa$-$squared on key FRET quantities. For example, in Figure \ref{fig-panel-fret-behaviour} (right) we presented the dependence of the FRET rate and F\"{o}rster radius with kappa$-$squared as derived from Eq. \ref{eq-kfret} and \ref{eq-r0}, respectively. Note that a semi$-$log plot is used to better visualization of FRET rate. The $k_{FRET}$ shows a linear increase with kappa$-$squared whereas $R_{0}$ has a non$-$linear variation with this parameter (see Eq. \ref{eq-r0}). These figures showed the importance of the orientation factor in the effective energy transfer by the FRET mechanism. The two commonly assumed values for kappa$-$squared are highlighted in the Figure \ref{fig-panel-fret-behaviour} (right).

\section{Conclusions}
FRET parameters play a fundamental role in the development and optimization of new technological applications. Thus, FRET$-$Calc is a robust tool developed to quickly obtain the effective refractive index, overlap integral, F\"{o}rster radius, FRET efficiency and FRET rate from experimental data. In this work the theoretical background and applicability of FRET$-$Calc (software and web server) were demonstrated in detail. FRET$-$Calc provides the researcher with parameters of the FRET process through a user$-$friendly interface, which also makes it suitable for teaching purposes. The FRET$-$Calc software and web server has free license to use and will be updated with new features. Furthermore, FRET-Calc is the first in a series of programs from our group aimed at analyzing experimental data oriented to the field of organic optoelectronics applications.

\section*{Declaration of Competing Interest}
\noindent The authors declare that they have no known competing financial interests or personal relationships that could have appeared to influence the work reported in this paper.

\section*{Acknowledgments}
\noindent The authors acknowledge financial support from CNPq (grant 381113/2021$-$3) and LCNano/SisNANO 2.0 (grant 442591/2019$-$5). L.B. (grant E$-$26/202.091/2022 process 277806), O.M. (grant E$-$26/200.729/2023 process 285493)  and G.C. (grant E$-$26/200.627/2022 and E$-$26/210.391/2022 process 271814) are greatfully for financial support from FAPERJ. The authors also acknowledge the computational support of N\'{u}cleo Avan\c{c}ado de Computa\c{c}\~{a}o de Alto Desempenho (NACAD/COPPE/UFRJ), Sistema Nacional de Processamento de Alto Desempenho (SINAPAD) and Centro Nacional de Processamento de Alto Desempenho em S\~{a}o Paulo (CENAPAD$-$SP) and technical support of SMMOL$-$solutions in functionalyzed materials.

\section{Data availability}
\noindent Data will be made available on request.

\bibliographystyle{elsarticle-num}
\bibliography{fret-bib}

\end{document}